\documentclass{aa}  

\usepackage{graphicx}
\usepackage{txfonts}
\usepackage{natbib,twoopt}

\usepackage[breaklinks=true]{hyperref} 
\bibpunct{(}{)}{;}{a}{}{,} 

\newcommand{\subrm}[2]{#1_{\rm #2}}

\begin{document} 
   \title{Can brown dwarfs survive on close orbits around convective stars?}
   \author{C. Damiani
          \inst{1}
          \and
          R. F. Díaz \inst{2}
          }
   \institute{Universit{\'e} Paris-Sud, CNRS, Institut d'Astrophysique Spatiale, UMR8617, 91405 Orsay Cedex, France\\
              \email{cilia.damiani@ias.u-psud.fr}
         \and
            Observatoire astronomique de l'Universit\'e de Gen\`eve, 
                   51 ch. des Maillettes, CH-1290 Versoix, Switzerland \\
             \email{rodrigo.riaz@unige.ch}
             }

   \date{Received 3 August 2015/ Accepted 23 February 2016}
  \abstract
   {Brown dwarfs straddle the mass range transition from planetary to stellar objects. There is a relative paucity of brown dwarfs companions around FGKM stars compared to exoplanets for orbital periods less than a few years, but most of the short-period brown dwarf companions fully characterised by transits and radial velocities are found around F-type stars. }
   {We examine the hypothesis that brown dwarf companions could not survive on close orbit around stars with important convective envelopes because the tides and angular momentum loss through magnetic breaking should lead to a rapid orbital decay and quick engulfment of the companion.}
   {We use a classical Skumanich-type braking law, and constant time-lag tidal theory to assess the characteristic timescale for orbital decay for the brown dwarf mass range as a function of the host properties. }
   {We find that F-type stars may host massive companions for a significantly longer time than G-type stars for a given orbital period, which may explain the paucity of G-type hosts for brown dwarfs with orbital period less than 5 days. On the other hand, we show that the small radius of early M-type stars contributes to orbital decay timescales that are only half those of F-type stars, despite their more efficient tidal dissipation and magnetic braking. For fully convective later type M-dwarfs, orbital decay timescales could be orders of magnitude greater than for F-type stars. Moreover, we find that for a wide range of values of tidal dissipation efficiency and magnetic braking, it is safe to assume that orbital decay for massive companions can be neglected for orbital periods greater than 10 days.}
   {For orbital periods greater than 10 days, brown dwarf occurrence should largely be unaffected by tidal decay, whatever the mass of the host. On closer orbital periods, the rapid engulfment of massive companions could explain the lack of G and K-type hosts in the sample of known systems with transiting brown dwarfs. However, the paucity of M-type hosts can not be an effect of tidal decay alone, but may be the result of a selection effect in the sample and/or the formation mechanism.}


   \maketitle
%

\section{Introduction}
Brown dwarfs are sub-stellar objects that occupy the mass range between the heaviest gas giants planets and the lowest mass stars, usually considered to be between 13 and 80 Jupiter masses. The higher end of this mass interval can be well defined by a relevant physical phenomenon, namely having insufficient mass to sustain hydrogen fusion reactions in their core. This clearly distinguishes two different kinds of objects: those able to be in nuclear equilibrium for most of their lifetime, defined as stars, and those that lack significant support against gravitational contraction, defined as brown dwarfs \citep{Burrows1997}. The lower bound mass for the definition of a brown dwarf has often been taken to be the deuterium-burning limit but such a distinction is currently debated, since deuterium burning has a negligible impact on stellar structure and evolution \citep{Chabrier2000}. Based on observations, it may be more relevant to distinguish brown dwarfs and giant planets based on their dominant formation mechanisms, which allows a mass overlap between these two populations \citep{Luhman2012,Chabrier2014}. Of course such a definition is not without its challenges, since it is not straightforward to infer the formation of a particular object based on its observable properties. Moreover, it is not yet clear wether brown dwarfs form like stars or if they are issued from a different formation scenario, that may or may not share common properties with the formation of planets \citep{Chabrier2014}. 

The statistics of brown dwarf and giant planet companions to stars potentially bears important information on the formation of this objects. In particular, an important result of the radial velocity surveys has been the identification the so-called "brown dwarf" desert, that is a lack of companions with a mass between 10 and 100 M$_{\rm J}$ relative to planetary or stellar companions within 3 AU around main-sequence FGKM stars \citep{Marcy2000,Grether2006,Sahlmann2011}. The frequency of companions decreases with increasing mass before the desert, but increases for heavier masses towards the stellar companion range. This may stem from qualitatively and quantitatively different formation mechanisms and help distinguishing brown dwarfs and giant planets \citep{Ma2014,Chabrier2014}. 

The results of transit surveys shed a different light on the brown dwarf desert.  However, one must bear in mind that transit surveys capable of detecting brown dwarf companion to main-sequence stars are putting their efforts mainly into exoplanets detection and characterisation. Nevertheless, close-in giant planets and brown dwarfs have approximately the same radius, therefore one could assume that they have the same transit detection bias. Recently, \citet{Csizmadia2015} computed the relative frequency of brown dwarfs and hot Jupiters based on \textit{CoRoT} \citep{Baglin2006} data, and found a brown dwarfs/hot Jupiters occurrence ratio of $\approx14$\%. They remark that it is inconsistent with the results of ground based transit surveys, for which they estimate an occurrence ratio of 0.05\%.They reckon that a major factor explaining the discrepancy may be that ground-based surveys are biased for the detection of transits smaller than 1\%. Likewise, \citet{2015arXiv151100643S} conducted a systematic survey of Kepler  \citep{2010Sci...327..977B} planetary candidates using radial velocities to determine their masses. They were able to derive the occurrence rate of brown dwarfs and giant planets on orbital periods up to 400 days. They find an occurrence rate of $0.29\pm0.17\%$ for brown dwarfs, which is 15 times smaller than the value of $4.6\pm0.6\%$ they derive for giant planets. They conclude that their values are in agreement with the brown dwarfs/hot Jupiters occurrence ratio derived by \citet{Csizmadia2015}, but they note that the brown dwarfs detected by \textit{CoRoT} have orbital periods of less than 10 days, while in the \textit{Kepler} sample, they have periods between 10 and 170 days. It should also be noted that their sample selects candidates with transit depth less than 3\%, which by construction rejects massive companions orbiting M stars. Therefore, to this day, there is no definitive census of the companion occurrence seen by transit surveys extending to the brown dwarf regime, but it appears that the occurrence rate of close-in brown dwarfs is smaller than that of close-in giant planets and possibly that the brown dwarf frequency decreases for shorter periods. This could either be the consequence of their formation mechanism or the result of strong tidal interactions that lead to the faster engulfment of more massive companions.

Already at the time of the first detection of a very massive transiting companion with a short orbital period \citep{Deleuil2008}, there were observational evidence that close-orbiting companions to F-type stars could be more massive than companions to lower mass host. A few years later, \citet{Bouchy2011} discussed again this trend and also put forward an interesting conjecture to try and explain the prevalence of F-type host amongst the transiting systems hosting a massive companion. They commented that tidal interactions alone do not necessarily result in the engulfment of the companion, because companions massive enough can reach instead a state of tidal equilibrium with synchronous orbital mean motion and stellar spin. But even in that case, magnetic braking in the central star would lead to a loss of angular momentum that is transferred to the orbit of the companion through tides and lead to orbital decay. It is known that early and mid-F-type dwarfs are typically rapid rotators, independently of their age, as a consequence of a small outer convective zone, weak stellar winds, and smaller losses of angular momentum. Thus they proposed that close-in massive planets and brown dwarfs could survive tidally induced orbital decay when orbiting early or mid F-type dwarfs, but be engulfed by G and late F-type dwarfs due to their more efficient magnetic braking.

Using a numerical model that includes tidal interactions, stellar evolution, magnetic braking and a consistent calculation of tidal dissipation $Q'$ by gravity waves, \citet{Guillot2014} have computed the survival time of companions to stars in the mass range 0.8 to 1.4~M$_\odot$, and with masses between 0.2 and 200~M$_{\rm J}$ on a initial orbital period of 3 days. They conclude that massive close-in planets and brown dwarfs are engulfed preferentially around G-dwarfs, but they also note that observations show an even stronger deficit of massive companions around G-dwarfs than found by their model. Recently, \citet{Mathis2015} computed the frequency-averaged tidal dissipation in the convective envelope of low-mass stars and found that it is very sensitive to the mass and aspect ratio of the radiative core. They estimated that, at a given stellar rotation rate, the frequency-averaged tidal dissipation efficiency in the convective envelope decreases with increasing mass on the main sequence for masses between 0.5 and 1.4M$_\odot$. Adding the tidal dissipation in the convective envelope to the one in the radiative zone treated by \citet{Guillot2014} could reconcile their results to observations. The quantitative effects of the evolution of the rotation of the star on the dissipation in the convective envelope remains however to be investigated.

In this paper, we put \citeauthor{Bouchy2011}'s hypothesis to the test and compare the dynamical evolution of massive companions around stars F and G stars, but we also extend the comparison to K and M hosts. We produce a general analytic formulation of the characteristic timescale for orbital decay while accounting for the combined effects of tides and magnetic braking. In Sec.~\ref{sec:tidaleq}, we recall how the inclusion of magnetic braking changes the configurations of tidal equilibrium, that is now a dynamical equilibrium state that evolves in time. In Sec.~\ref{sec:interplay} we use the equilibrium tide theory and a classical Skumanich-type braking law to evaluate the ranges of orbital periods and rotation periods that correspond to different orbital decay regimes, depending on the distance to the dynamical equilibrium state. We also provide quantitative estimate for the survival time of a massive companion around dwarfs with an outer convective envelope surrounding a radiative zone. In Sec.~\ref{sec:discuss}, we discuss the resulting observable effects on the relative frequency of massive companions around different types of host. Finally in Sec.~\ref{sec:conclusion}, we give our conclusions.	

\section{Tidal pseudo-equilibrium}\label{sec:tidaleq}
We consider a system formed by a star and a gravitationally bound companion of masses $M_\star$ and $M_c$, respectively, and radii $R_\star$ and $R_c$. They are considered as rigid bodies with moments of inertia about their rotation axis that are noted $C_\star$ and $C_c$ respectively. Both those moments can be written as $C = M (r_g R)^2$ where $r_g$ is the non-dimensional radius of gyration. The periodically varying potential experienced by both objects generates a tidal disturbance in the fluid. Regardless of the mechanism, dissipation of the tides is directly associated with the secular transfer of angular momentum between the spin and the orbit, as well as a loss of energy from the system.  In a closed system, the system evolves towards a minimum of energy while the total angular momentum $L$ remains constant. But the open lines of the magnetic field of the host can support a magnetised wind that efficiently extracts angular momentum from the star with a very low mass loss rate. In this case where neither the total energy nor the total angular momentum is conserved, the configuration corresponding to the minimum of energy evolves with time. Previous studies have shown the importance of magnetic braking for the dynamics of exoplanetary systems \citep{Bolmont2012, Ferraz2015}. But even without a detailed knowledge of the tidal dissipation mechanism or wind braking efficiency and reasoning only on the extrema of the total energy of the system as a function of the orbital elements, the outcome of tidal evolution under the constraint of angular momentum loss can be assessed. Following \citet{1980A&A....92..167H}, \citet{Damiani2015} have shown that the system evolves towards a minimum of energy that it is characterised by a circular and aligned orbit, as in the case of conserved angular momentum, but the synchronisation condition becomes 
\begin{eqnarray}
\omega=& n\\
\Omega=& \beta(t)\, n \label{quasicorot}
\end{eqnarray}
where $\omega$ and $\Omega$ are the angular velocity of the companion and the star respectively, $n$ is the mean orbital motion and $\beta(t)$ is a function of time defined as
\begin{equation}\label{eqbeta}
\beta(t) = 1 - \frac{{\rm d} L}{{\rm d} t} \left(C_\star \frac{{\rm d} \Omega}{{\rm d} t}\right)^{-1}.
\end{equation}

The parameter $\beta(t)$ can be seen as the ratio of the tidal torque to the total torque acting on the star. A value $\beta(t) \approx 1$ corresponds to the case where the total angular momentum of the system is approximately conserved and is equivalent to the case where magnetic braking is neglected. A value $\beta(t) \approx 0$ corresponds to the case where the total angular momentum loss of the system is the angular momentum loss of the star through its wind and the tidal torque acting on the star is negligible. The tidal torque can only spin up the star when $\Omega < n$. As shown in  \citet{Damiani2015}, the dynamical equilibrium state is possible only when $\Omega<n$. Provided that $\beta < 1$, at a given time $t$, \citet{Damiani2015} showed that the dynamical equilibrium can be reached only if the total angular momentum exceeds a critical value $L_{\rm crit}(t)$ given by
\begin{equation} \label{eq:Lcrit}
L_{\rm crit}(t)= 4 {\left( \frac{G^2}{3^3}\frac{M_\star^3 M_c^3}{M_\star +M_c } \Bigl( \beta(t) C_\star + C_c \Bigr)\right)}^{1/4}. 
\end{equation}

This value depends on $\beta(t)$, which is time dependent, and it requires $\Omega < n$. Thus as the system evolves the conditions for the existence of equilibrium also change, but the dynamical equilibrium state can be reached when $0 < \beta(t) < 1$ and $L(t) > L_{\rm crit}(t)$. In other words, the orbital decay is significantly affected by magnetic braking as long as the system can maintain $L(t) > L_{\rm crit}(t)$ and enter into the pseudo-equilibrium state. This stationary state could be stable or unstable depending on the sign of the second partial derivative of the energy in this state. It can be shown that the equilibrium will be stable if the orbital angular momentum $h$ satisfies
\begin{equation}\label{eq:stabcond}
h > (4-\beta(t))(C_c + C_\star)n.
\end{equation}
Let us consider the critical angular momentum in the absence of magnetic braking, i.e. when $\beta=1$. It is expressed as
\begin{equation} \label{eq:Lcrit0}
L_{\rm crit_0}= 4 {\left( \frac{G^2}{3^3}\frac{M_\star^3 M_c^3}{M_\star +M_c } \Bigl( C_\star + C_c \Bigr)\right)}^{1/4}. 
\end{equation}
At $L=L_{\rm crit_0}$, the unique mean motion corresponding to co-rotation is
\begin{equation} \label{eq:ncrit0}
n_{\rm crit_0}= \left( \frac{G^2}{3^3}\frac{M_\star^3 M_c^3}{M_\star +M_c } \right)^{1/4}\Bigl( C_\star + C_c \Bigr)^{-3/4}. 
\end{equation}

The values of $L_{\rm crit_0}$ and $n_{\rm crit_0}$ only depend on the masses and radii of the star and the companion, and there is no need to know the value of $\beta$ to compute them. Moreover, in the case where magnetic braking is present, it can be shown that $n_s(t)$, defined as the maximum orbital frequency allowing the existence of a pseudo-stable state is given by
\begin{equation}
n_s(t)=\left(\frac{3}{4-\beta(t)}\right)^{3/4}n_{\rm crit_0}.
\end{equation}
Since the equilibrium state can be reached only if $0 < \beta(t) < 1$, $n_s(t)$ always verifies
\begin{equation}
n_{\rm crit_0} \left(\frac{3}{4}\right)^{3/4}< n_s(t)< n_{\rm crit_0}.
\end{equation}
Similarly, it can be shown that the corresponding condition for the existence of a pseudo-stable state in terms of total angular momentum satisfies $L_{\rm crit}(t) < L_{\rm s}(t) < L_{\rm crit_0}$. Thus as long as the orbital mean motion $n \lesssim 0.8\, n_{\rm crit_0}$  and $L > L_{\rm crit_0}$, the system can reach a pseudo-stable equilibrium. In this way, provided that the orbital motion is prograde, the companion may not be directly engulfed into the star as long as  $\subrm{P}{orb}\gtrsim 1.25 P_{\rm crit_0}$, where $P_{\rm crit_0} = 2\pi/n_{\rm crit_0}$. Since $P_{\rm crit_0}$ depends only on the masses and radii of the star and the companion, the closest orbital period allowed, purely based on stability principles, can be computed even without a detail knowledge of the tidal dissipation efficiency or the wind braking.
   \begin{table}
      \caption[]{Host properties used to plot Fig.~\ref{fig:pcrit}. Spectral type and mass from \citet{Allen}, radius and gyration radius taken from CESAM stellar models \citep{1997A&AS..124..597M,2008Ap&SS.316...61M} at the ZAMS.}
         \label{tab:hosts}
         \centering
         \begin{tabular}{llll}
            \hline
            Spectral Type      &  $M_\star [\mathrm{M}_\odot] $ & $R_\star [\mathrm{R}_\odot]$  & $r_g$ \\
            \hline
            F0 &1.6 & 1.44 & 0.25\\ 
            F5 & 1.4 & 1.35 & 0.26\\
            G5 & 0.92 & 0.81 & 0.38\\
            K5 & 0.67 & 0.61 & 0.44\\
            M0 & 0.51 & 0.45 & 0.50\\
            M5 & 0.21 & 0.22 & 0.56 \\
            \hline
           \end{tabular}%
          \end{table}%
   \begin{figure}
   \centering
   \includegraphics[width=1.0\columnwidth]{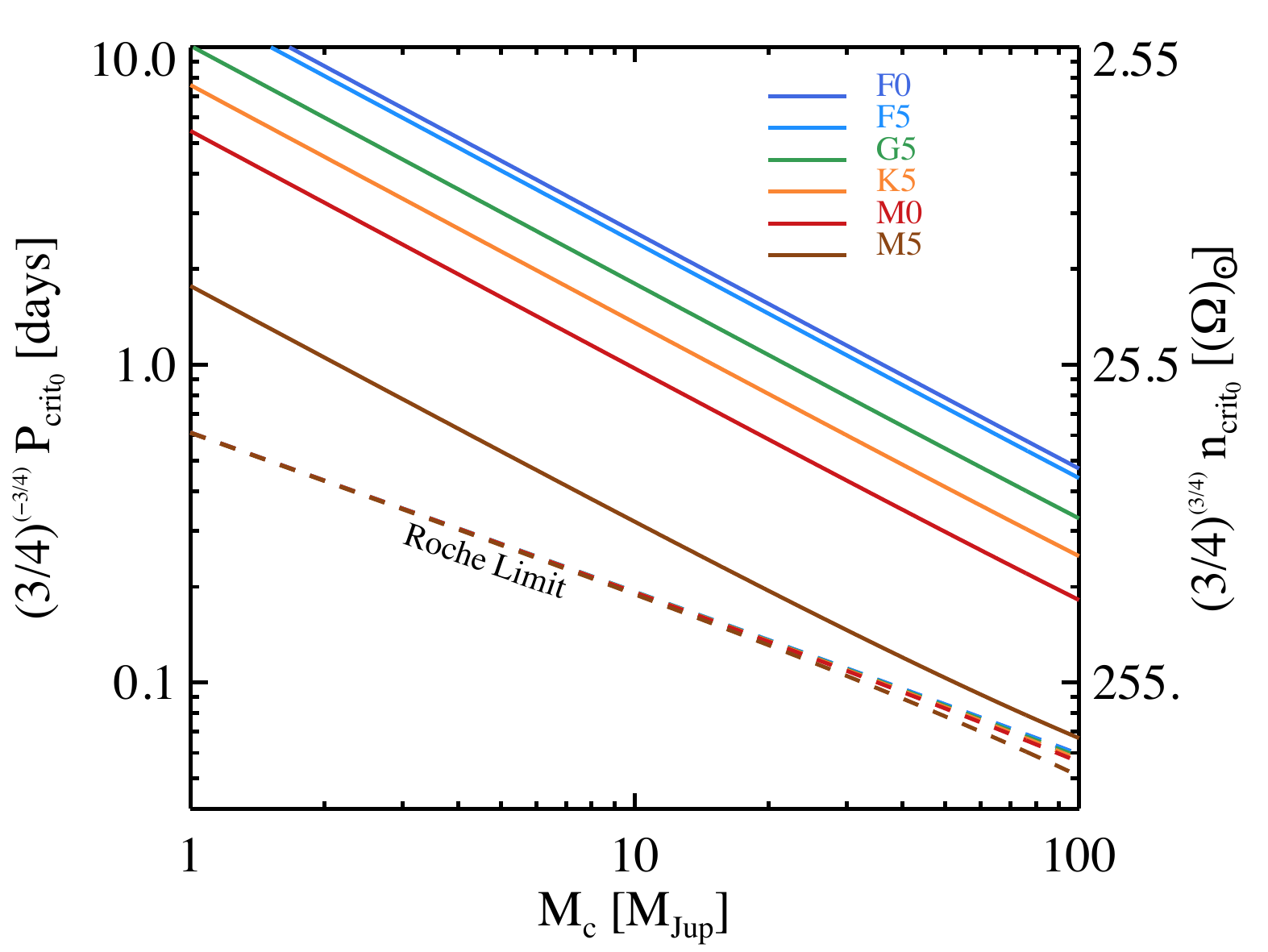}
      \caption{Upper bound on the closest orbital period allowing the existence of a pseudo-stable equilibrium state (provided that $\Omega < n$) as a function of the mass of the companion for different hosts (solid lines). The colour of the line corresponds to the spectral type of the host, as indicated on the plot. The orbital period corresponding to the Roche limit is given as dashed-line, with the same colour convention.}%
         \label{fig:pcrit}
   \end{figure}
 We give this quantity as a function of the companion in Fig.~\ref{fig:pcrit} for different host masses. The properties of the host stars are given in Table.~\ref{tab:hosts}. The Roche limit, where we suppose that the companion may be disrupted by tides is also plotted in this figure by computing the orbital period correspond to the semi-major axis $a_{\rm R}$ defined as: 
\begin{equation}
a_{\rm R} = 2.422 R_c \left(\frac{M_\star}{M_c}\right)^{1/3}
\end{equation}
taking $R_c = 1.2 R_{\rm J}$ for the whole mass range.

Firstly, we see that the Roche limit is closer to the star than the closest stable orbit. Thus the survival of the companion is not threaten by tidal disruption before it can enter the stationary state. Secondly, we see that for a given companion mass, the closest stable orbit gets closer with decreasing stellar mass. Thus convective star should be able to retain massive companion on closer orbit than the more massive hosts. Thirdly, for a given host mass, the closest stable orbit gets closer to the host with increasing companion mass. Thus more massive companions should be able to survive on orbits where planets are found. The critical factor differentiating massive companions and Jupiter-like planet will be the characteristic time-scale of orbital decay through the interplay between tidal interaction and magnetic braking. For a given host mass, rotation frequency and orbital period, more massive companions have greater total angular momentum. Through Eq.~\eqref{eq:Lcrit}, this makes them more likely to enter the stationary state. But the value of their tidal torque will also be greater than for less massive planets, which increases the value of $\beta$ and, through Eq.~\eqref{eq:stabcond}, sets a higher value on the required ratio of angular momentum to ensure their stability. It is thus necessary to describe more accurately the evolution of the tidal torque and the wind torque in time, and the resulting evolution of the orbital parameters.	

\section{Interplay between the tidal torque and the wind torque}\label{sec:interplay}
   \begin{figure}
    \centering
   \includegraphics[width=0.9\columnwidth]{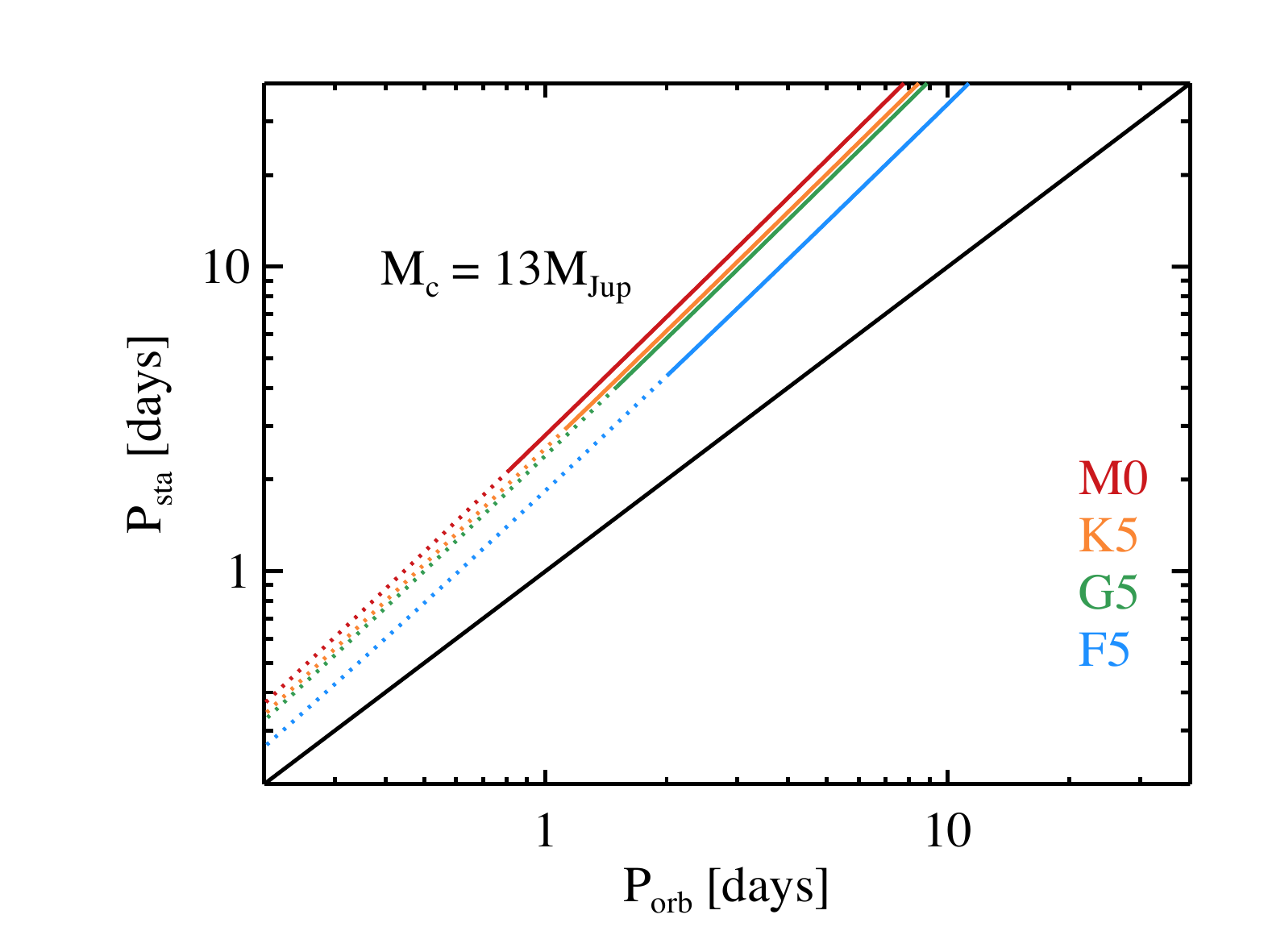}\\
   \includegraphics[width=0.9\columnwidth]{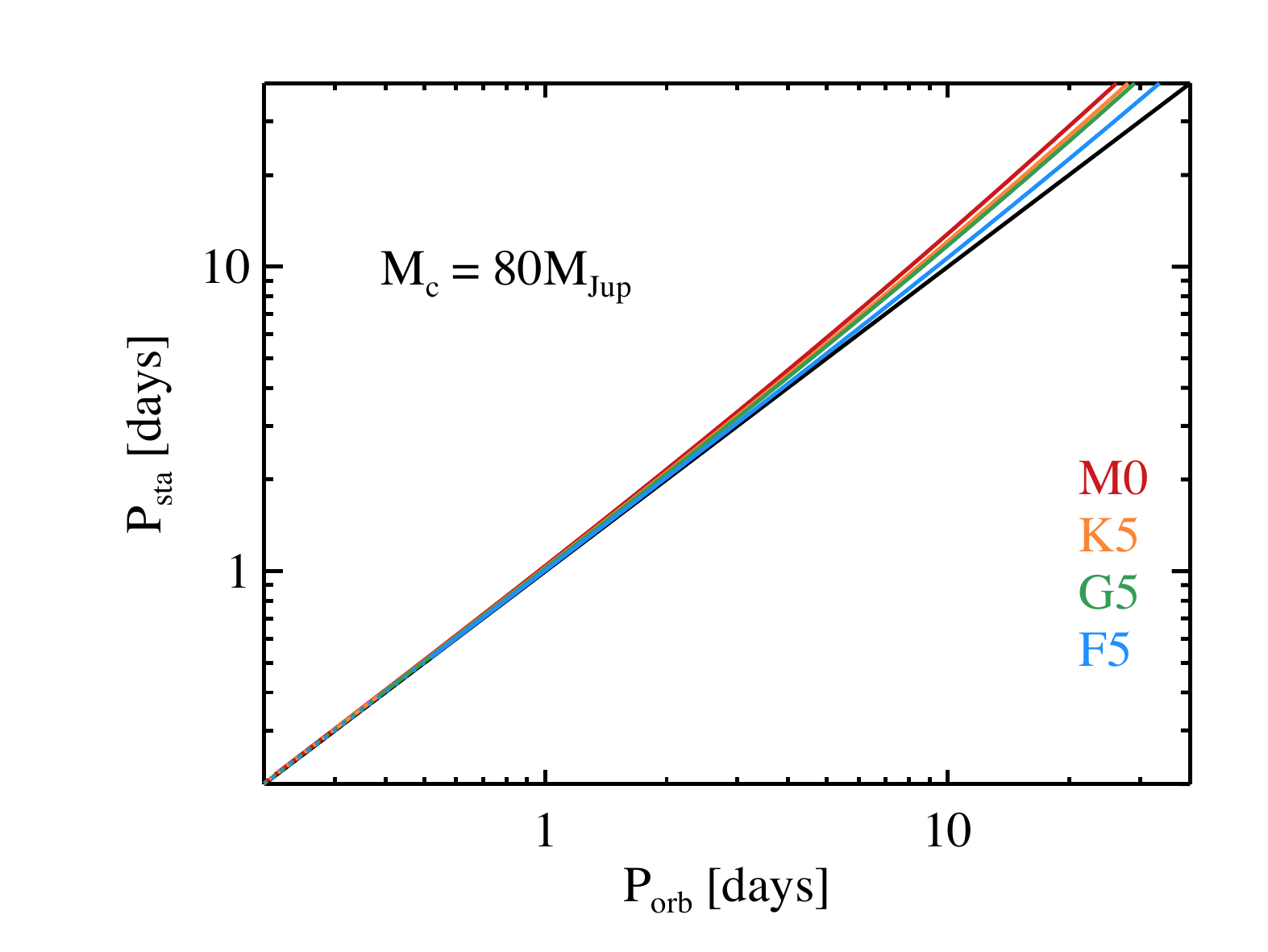}\\
     \caption{Stellar rotation period corresponding to torque balance as a function of the orbital period in days for companions of masses $M_{\rm c}=$~13 (top) and 80 (bottom) Jupiter masses and hosts of different masses (cf.~Tab.~\ref{tab:hosts}) shown in different colours as indicated on the plots. The solid portion of the lines correspond to orbits where the stationary state is stable, and the dotted portion shows the unstable stationary state (when $\subrm{P}{orb}\lesssim 1.25 P_{\rm crit_0}$). The solid black line on each plot represents the synchronisation between the orbital period and the stellar spin period.}%
    \label{fig:psta}
   \end{figure}
By examining the total mechanical energy of the system as a function of the orbital elements, we have characterised the orbital configuration corresponding to the minimum of energy. The mechanical energy of the system is not conserved because dissipative fluid processes convert the mechanical energy of the tidal torque into heat, but also because the stellar wind removes kinetic energy from the system. To assess the rate of change of the orbital elements, we must assume some form of tidal dissipation and magnetic braking. Following \citet{Damiani2015}, we use a formulation based on \citet{Barker2009}, obtained in the framework of the equilibrium tide assuming a constant $Q'$. Adopting a constant $Q'$ implies that the time lag between the maximum of the tidal potential and the tidal bulge in each body scales with the orbital period and that the relevant tidal frequency is the orbital frequency. This may not give identical numerical factors than other formulations of tidal friction, but given our lack of knowledge of the actual values of $Q'$ and its dependence on tidal frequency, this approximation remains reasonable (see \citet{Mathis2013} and \citet{Ogilvie2014} for detailed reviews on tidal dissipation). We use a Skumanich-type law for magnetic braking with a torque of magnitude $\Gamma_{\rm mb} = -\alpha_{\rm mb} C_\star \Omega^3$, where the value of $\alpha_{\rm mb}$ is estimated from observed rotation periods of stars of different ages. This formulation overlooks the complex physics of stellar winds \citep{Matt2012,Reville2015}, but for the sake of generality, we will assume that magnetic braking depends only on the mass, radius and rotation rate of the star \citep[as in][]{Kawaler1988}. Following \citet{Dobbs-Dixon2004}, we take $\subrm{\alpha}{mb} = 1.5 \times 10^{-14} \gamma$~yr where $\gamma = 0.1$ for F stars and $\gamma=1.0$  for latter type stars. To account for a possible relationship between the tidal dissipation efficiency and the extent of the convective envelope, we use a value of $Q^\prime= 10^8$ for F stars, $Q^\prime= 10^6$ for latter type stars \citep{Mathis2015,Mathis2015sf2a,Ogilvie2007}. Those values are somewhat arbitrary, but we do not attempt to precisely estimate an absolute time for the survival of a companion on close orbit, but rather we aim at comparing the survival time of companions of different mass around different kind of stars. This qualitative parametrisation should be realistic for stars with an outer convective envelope surrounding a radiative zone, because we expect their dynamos and tidal dissipation mechanisms to be similar.  We assume that the companion is quickly synchronised with the orbit, so that the evolution of the system only depends on the tides raised in the star and its magnetic braking.

The temporal evolution of the stellar spin frequency and the orbital mean motion follows the following set of dimensionless equations:
\begin{align}
\frac{{\rm d}\tilde{\Omega}}{{\rm d}\tilde{t}} &= \tilde{n}^4 \left(1-\frac{\tilde{\Omega}}{\tilde{n}}\right) - A \tilde{\Omega}^3,\label{omegadot}\\
\frac{{\rm d}\tilde{n}}{{\rm d}\tilde{t}} &= 3 \tilde{n}^{16/3} \left(1-\frac{\tilde{\Omega}}{\tilde{n}}\right),\label{ndot}
\end{align}
where $\tilde{\Omega}$ and $\tilde{n}$ are dimensionless variables that are related to the ones previously defined by the following relationships: 
\begin{align}
\tilde{n} = \frac{n}{\subrm{n}{crit_0}} 3^{-3/4}, & \qquad
\tilde{\Omega} = \frac{\Omega}{\subrm{n}{crit_0}} 3^{-3/4},
\end{align}
and $A$ is a non-dimensional constant defined as
\begin{equation}\label{eq:a}
A=\frac{2}{3^{9/4}}\subrm{\alpha}{mb} Q^\prime\subrm{n}{crit_0}\frac{\subrm{M}{\star}}{\subrm{M}{c}} \subrm{r}{g}^{5}  \left(\frac{\subrm{M}{c}}{\subrm{M}{\star}+\subrm{M}{c}}\right)^{-5/2} .
\end{equation}
The stationary state, i.e when the torque exerted on the star by the wind is balanced by the tidal torque, is equivalent to $\dot{\Omega} =0$. According to Eq.~\ref{omegadot}, this means: 
\begin{equation}\label{cubic}
\tilde{\Omega}^3 + \frac{\tilde{n}^3}{A} \tilde{\Omega} -  \frac{\tilde{n}^4}{A} = 0. 
\end{equation}
The discriminant of this cubic equation in $\tilde{\Omega}$ is always negative when $\tilde{n}>0$, thus for each positive $\tilde{n}$ there is one real value $\tilde{\Omega}_{\rm sta}$ corresponding to the torque balance. Using Cardano\rq{}s method, the real root of Eq.\ref{cubic} can be written as:
\begin{equation}\label{eqstat}
\tilde{\Omega}_{\rm sta} = \tilde{n} \, \sqrt[3]{\frac{\tilde{n}}{2A}} \left(\sqrt[3]{1+\sqrt{1+\frac{4\tilde{n}}{27A}}} +\sqrt[3]{1-\sqrt{1+\frac{4\tilde{n}}{27A}}  }\right). 
\end{equation}
The corresponding stationary period $\subrm{P}{sta}$ is shown in Fig.~\ref{fig:psta} as a function of the orbital period for several masses of host and companion. 

Let us note that when the system has a spin period $\subrm{P}{spin}$ such as $\subrm{P}{spin} < \subrm{P}{orb}$, both the tidal torque and the wind torque act to spin down the star and, the tides push the companion outward. In this case, tidal interaction is not threatening the survival of the companion. But as the star keeps loosing angular momentum, it will eventually reach a state where $\subrm{P}{\rm spin} \geq\subrm{P}{orb}$ and tidal evolution may lead to orbital decay and threaten the survival of the companion. We will focus our study on this kind of orbital configuration.
\subsection{Timescales of evolution}

For a circular and aligned system, and assuming that the companion is synchronised with the orbit, the rate of variation of the total angular momentum of the system is
\begin{equation}\label{eq:dotL}
\frac{{\rm d} L}{{\rm d} t} = \frac{{\rm d} h}{{\rm d }t} + C_\star \frac{{\rm d} \Omega}{{\rm d }t} = -\alpha_{\rm mb} C_\star \Omega^3.
\end{equation}
When the wind torque is much smaller than the tidal torque acting on the star (i.e. when $\subrm{P}{\rm spin} > \subrm{P}{sta}$, see Fig.~\ref{fig:psta}), we can assume that the total angular momentum of the system is approximately conserved so that $\dot{L}\approx0$ and the in-spiral time of the orbit $\tau_a$ can be written \citep{Barker2009}
\begin{align}
\tau_a & \equiv -\frac{2}{13}\frac{a}{\dot{a}}\nonumber\\
& \approx12.0 {\rm Myr} \left(\frac{Q'}{10^6}\right) \left(\frac{M_\star}{M_\odot}\right)^{\frac{8}{3}} \left(\frac{M_c}{M_{\rm J}}\right)^{-1} \left(\frac{R_\star}{R_\odot}\right)^{-5} \left(\frac{\subrm{P}{orb}}{1 {\rm d}}\right)^{\frac{13}{3}} \left(1-\frac{\subrm{P}{orb}}{\subrm{P}{\rm spin}}\right)^{-1}.\label{eq:tauatides}
\end{align}

If the orbit is inside co-rotation, angular momentum is transferred from the orbit to the spin of the star at a rate that is set by the tidal torque alone. The spin-up time of the star is then
\begin{equation}\label{eq:tauomtides}
\tau_{\Omega}  \approx \frac{13}{2} \frac{C_\star \Omega}{h} \tau_a
\end{equation}

On the other hand, when the wind torque is much stronger than the tidal torque acting on the star (i.e. when $\subrm{P}{orb} \leq\subrm{P}{\rm spin} \leq \subrm{P}{sta}$), we can assume that
\begin{equation}\label{eq:tauomwind}
\frac{{\rm d}\Omega}{{\rm d} t}  \approx -\alpha_{\rm mb} \Omega^3.
\end{equation}
and the spin-down of the star takes place on the characteristic time of the magnetic braking
\begin{equation}
\tau_{\Omega}  \approx \frac{1}{\alpha_{\rm mb}} \frac{1}{\Omega^2},
\end{equation}
then by Eq.~\eqref{eq:dotL} we have 
\begin{equation}
\frac{{\rm d} h}{{\rm d }t} \approx 0
\end{equation}
so we can assume that $h\approx$~const, so that for a circular orbit, the orbital period remains constant.

When the system is close to the pseudo-stable stationary state where the wind torque is opposite in sign and equal in magnitude to the tidal torque (i.e when $\subrm{P}{\rm spin} \sim \subrm{P}{sta}$), we have $\dot{\Omega}\approx 0$ and $\Omega\approx \subrm{\Omega}{sta}$. From Eq.~\eqref{eq:dotL} we get
\begin{equation}
\frac{{\rm d} h}{{\rm d }t} \approx -\alpha_{\rm mb} C_\star \subrm{\Omega}{sta}^3.
\end{equation}
Since the orbital angular momentum is given by
\begin{equation}
h^2=G\frac{M_c^2 M_\star^2}{(M_c +M_\star)}a ,
\end{equation}
we have 
\begin{equation}
\frac{\dot{h}}{h} = \frac{1}{2}\frac{\dot{a}}{a}
\end{equation}
so that 
\begin{equation}\label{eq:tauastat}
\tau_a \approx \frac{1}{13}\frac{h}{\alpha_{\rm mb} C_\star \subrm{\Omega}{sta}^3}.
\end{equation}
In this way we see that when the system is close to the  pseudo-stable stationary state, the rate of orbital decay depends explicitly on magnetic braking, but it has also a non linear dependance with the efficiency of tidal friction through the value of $\subrm{\Omega}{sta}$. 

As the tides transfer angular momentum from the orbit to the star, any system that would enter the pseudo-stable stationary state would eventually become unstable by exhaustion of orbital angular momentum. Whenever $\subrm{P}{orb}\lesssim P_{\rm crit_0}$ the stationary state is unstable, and the in-spiral characteristic time is not given by Eq.~\eqref{eq:tauastat} because the system moves away from the torque balance as soon as it reaches it, and $\dot{\Omega} \neq 0$. In this case, the relevant timescale for orbital decay is given by Eq.~\eqref{eq:tauatides}. It is easy to show that then $\tau_\Omega \geq \tau_a$, which neglects the spin-down effects of magnetic braking. Indeed, by Eq.~\eqref{eq:stabcond}, we get that for unstable systems, the orbital angular momentum is at most four times the total spin angular momentum. The moment of inertia of the companion being small compared to the host, it is safe to consider that when the stationary state is unstable, the ratio of stellar spin to orbital angular momentum is large enough to let the tidal spin-up time of the star be larger than the in-spiral time by Eq.~\eqref{eq:tauomtides}. Thus even for rapidly rotating hosts, when the stationary state is unstable, the relevant timescale for orbital decay is given by Eq.~\eqref{eq:tauatides}. Close to the stationary state, the characteristic time-scale for orbital decay is thus different wether the stationary is pseudo-stable or not. The former depends non-linearly on magnetic braking and tidal dissipation efficiency, and the later is a linear function of tidal dissipation efficiency alone.

\begin{figure}
    \centering
   \includegraphics[width=1.0\columnwidth]{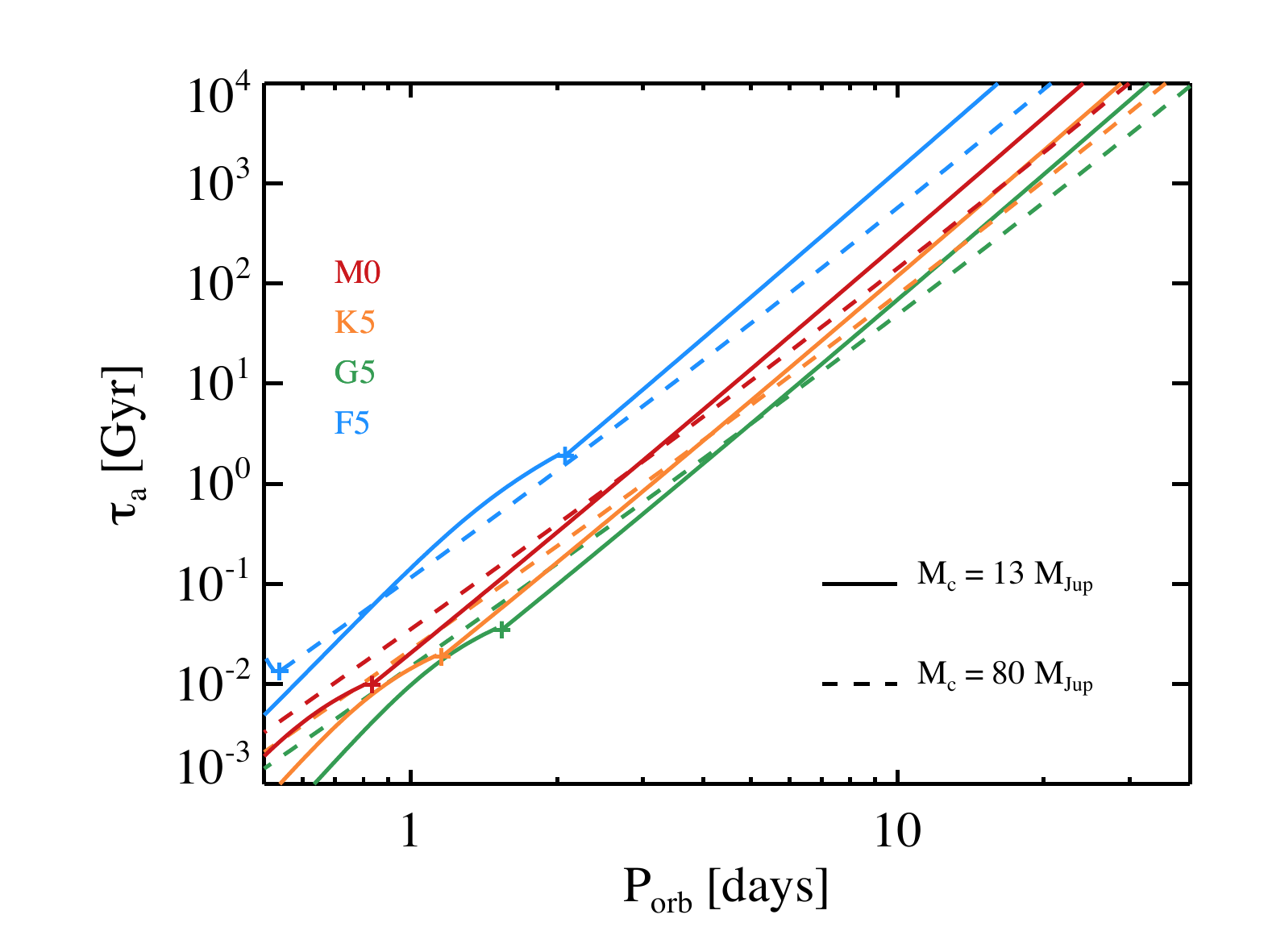}
     \caption{Typical time-scale for tidal decay $\tau_a$ as a function of the orbital period for different masses of host and companion as indicated on the figure. The cross symbol outlines the transition where the stationary state can no longer be stable and the in-spiral time stops being set by the magnetic braking of the star. See text for details.}%
         \label{fig:taua}
   \end{figure}

In Fig.~\ref{fig:taua}, we give characteristic in-spiral times as a function of the orbital period for different host and companion masses. When the pseudo-stable stationary state is stable, $\tau_a$ is computed with Eq.~\eqref{eq:tauastat} taking  $\subrm{P}{\rm spin} =\subrm{P}{sta}$. When $\subrm{P}{orb}< \subrm{P}{c_0}$, the stationary state can no longer be stable therefore, we compute $\tau_a$ using Eq.~\eqref{eq:tauatides}. We must assume a different value for the spin of the star, since the system is no longer in the stationary state. As discussed above, in this case the evolution proceed as if the total angular momentum were conserved  (the characteristic timescale for the spin down of the star is much slower than the tidal spin-up). The rotation period of the star is a function of the (conserved) total angular momentum and the orbital period only.  To illustrate the transition between the two regimes, we have set the value of the total angular momentum in the unstable state to the one it had at the end of the pseudo-stable stationary state. The transition between the two regimes is materialised by a cross symbol in the curves. There, the stationary state can no longer be stable and the in-spiral time stops being set by the magnetic braking of the star.
 
We see that whatever the orbital period, the in-spiral time of companions more massive than 13~M$_{\rm Jup}$ is about one order of magnitude greater around F-type stars, than around G and K-type stars. This provides a quantitative measure of the difference between G-K and F stars first evoked by \citet{Bouchy2011}, and agrees with the sophisticated simulations of \citet{Guillot2014}. For the early M-type stars, the in-spiral time is about half that of F-type stars, despite their being almost entirely convective. Indeed, the moment of inertia of an F-type star is typically two orders of magnitude greater than the one of an M-type star, mainly due their difference in radius. On the other hand the orbital angular momentum scales as the inverse square root of the host mass. Thus for given companion mass and orbital distance, the orbital angular momentum around F-type star is just a few times greater than when orbiting an M-type star. The ratio of orbital to spin angular momentum can thus be one to two orders of magnitude greater for an F-type star than for an M-type star. As can be seen form Eq.~\eqref{eq:tauastat}, this partly compensates for the difference in their magnetic braking efficiency.
   \begin{figure}
   \centering
   \includegraphics[width=0.9\columnwidth]{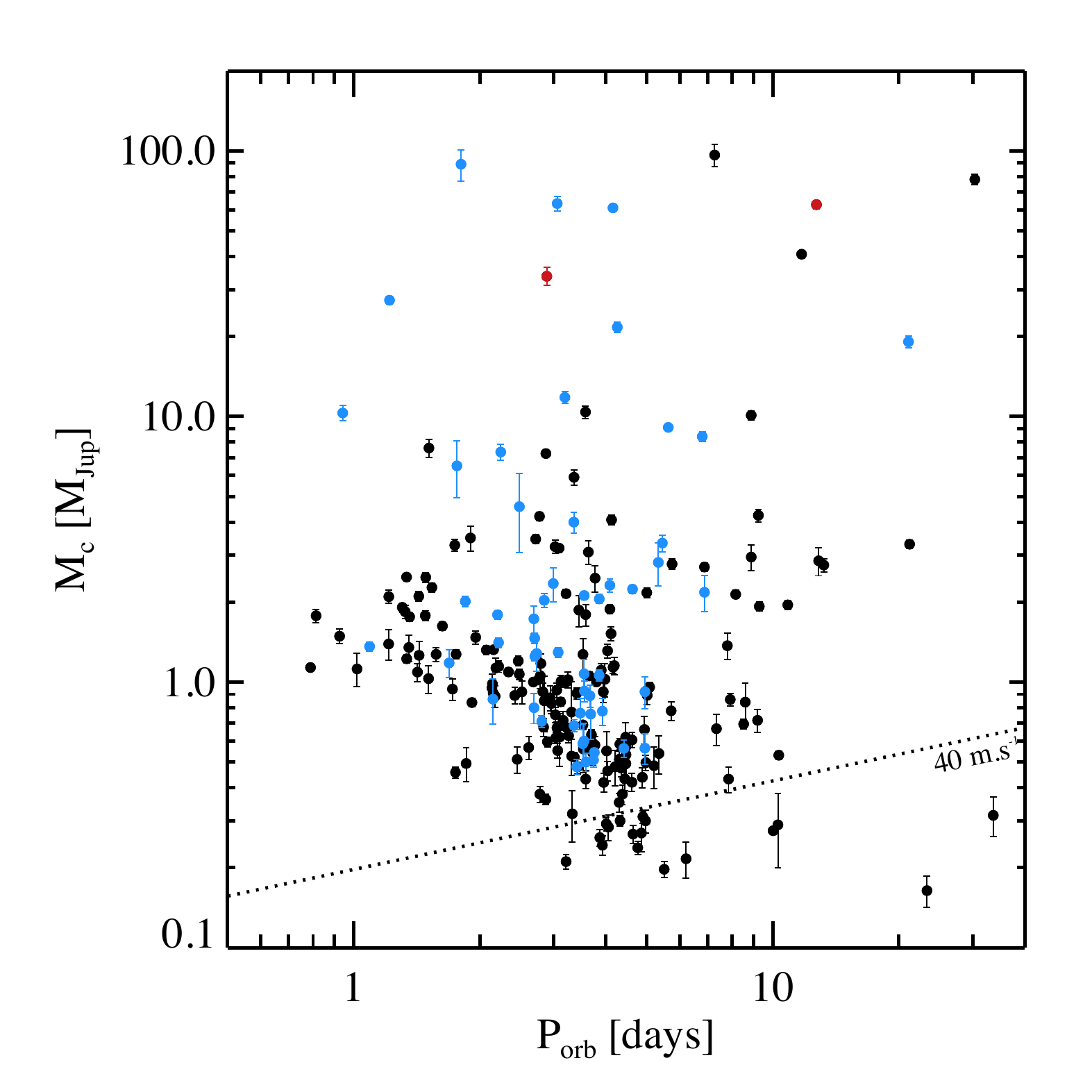}
      \caption{Mass versus period of well characterised transiting companions with $M_c \geq 0.1$ and $\subrm{P}{orb}\leq 40$~days. The colour of the symbol is blue if the host star has $T_{\rm eff} \geq 6200$~K, red if $T_{\rm eff} \leq 3300$~K and black otherwise. The detection limit well within the reach of typical radial velocity surveys capabilities is given by dotted line.}%
         \label{fig:massvsporb}
   \end{figure}

\section{Discussion}\label{sec:discuss}
   \begin{figure*}
   \centering
   \includegraphics[width=1.0\textwidth]{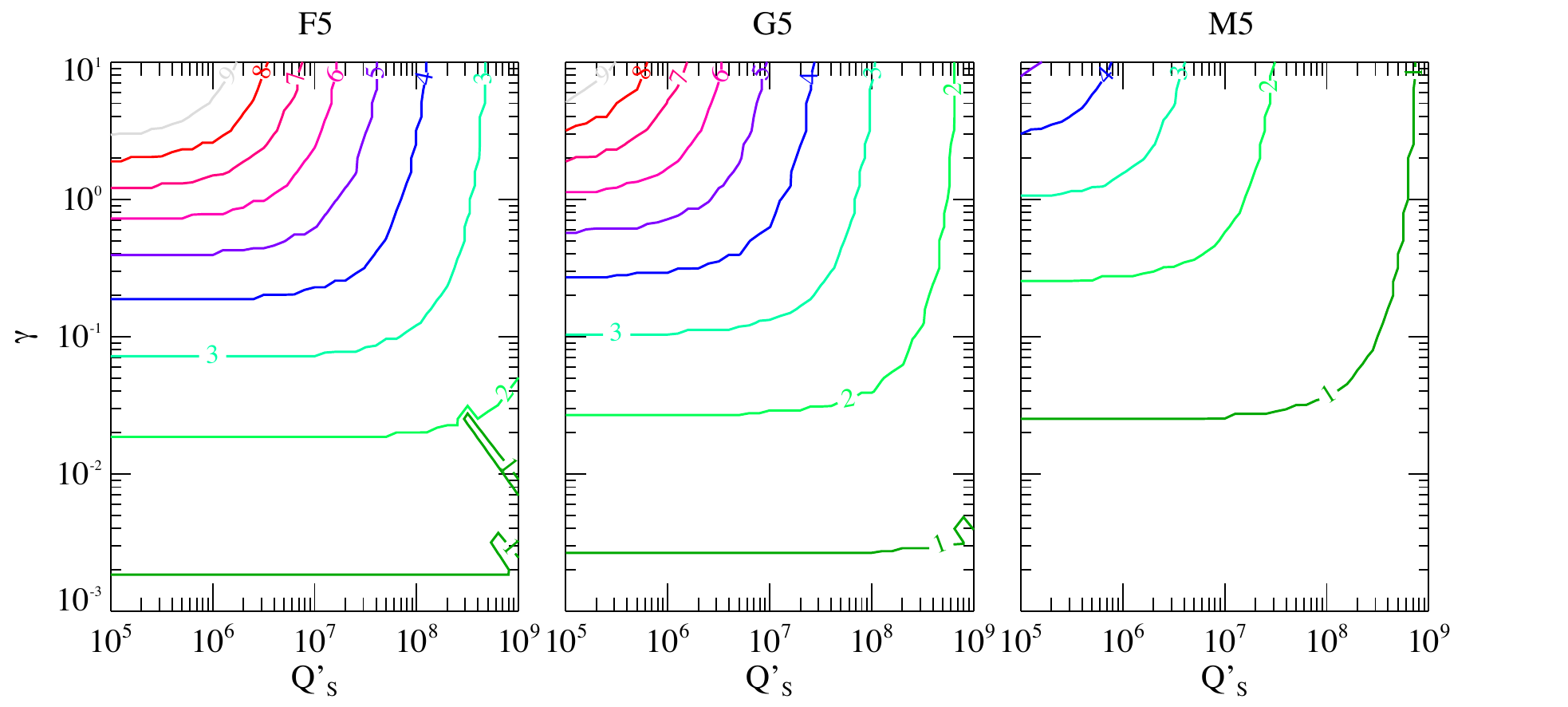}
      \caption{Contours of the minimum orbital period of a 80 Jupiter mass companion for which $\tau_a\geq 5$~Gyrs as a function of the tidal quality factor in the star $Q'_s$ and the coefficient $\gamma= \subrm{\alpha}{mb}/1.5 \times 10^{-14}$~yr of the magnetic braking law . Different spectral type of the host are sorted in columns as labeled on the figure.}%
         \label{fig:perlim}
   \end{figure*}

 Assuming that the values we chose for the tidal dissipation and magnetic braking are reasonable, we find that for massive companions around G and K stars, the characteristic time-scale for orbital decay is shorter than $\sim$~10~Gyr only for orbital periods closer than about 5 days. On the other hand, they can survive over the same time-scale down to orbital periods of about 3 days when orbiting an F-type star,  and about 4 days  for early M-type host. As conjectured by \citeauthor{Bouchy2011}, this would indeed explain the prevalence of F-type host amongst the transiting systems hosting a massive companion on very close orbit. Besides, based on the same reasoning, it is also expected to see massive companions surviving on closer orbits around M-stars than around G-stars.

On Figure~\ref{fig:massvsporb}, we present the mass as a function of the orbital periods up to 40 days for systems with measured radial velocities and detected transits. Out of the 11 well characterised transiting companions more massive than $\sim13$~M$_{\rm J}$, 6 are orbiting stars with $T_{\rm eff} \geq 6200$~K. This number becomes to 5 out of 7 if we consider companions with orbital periods shorter than 10 days, and 5 out of 6 for orbital periods shorter than 5 days. The number of BD companions is small and robust statistics are difficult to establish, but it appears that observations are consistent with the idea that the combined effect of tidal dissipation and magnetic braking shape the distribution of massive companions on very short orbits, in agreement with \citeauthor{Bouchy2011}'s hypothesis.

They strictly discussed the qualitative difference in tidal effects expected between early or mid F-type on one hand and G and late F-type stars on the other hand. We have shown here that massive companions could also survive on a close-in orbit for a long time around early M-dwarfs, despite their efficient dissipation and braking. And indeed, we see that for masses greater than about 10~M$_{\rm J}$ and orbital periods shorter than 5 days, all the systems discovered so-far are found exclusively around F or M stars. The large preponderance of F-type host compared to M-type host may be readily attributed to observational biases against M-type host in the sample. First those stars are faint and difficult to follow-up with ground-based radial velocities but they also have a deeper transit that excludes them from exoplanet search programs, which is the main focus of short-period transit surveys. A systematic survey of the transiting short-period companions to M-type stars, whatever their mass range, could easily validate our theory.

Our estimation of the characteristic time-scale for orbital decay as a function of the orbital period is sensitive to the choice of parameters. Indeed, studies attempting to calibrate the stellar modified tidal quality factor $Q'$ with the observed distribution of orbital periods of exoplanets find a value somewhat greater than what is considered here \citep{Jackson2009, Hansen2010}. However they generally neglect the effect of magnetic braking and seem to be inconsistent with the circularisation of stellar binaries \citep{Ogilvie2007}. We cannot exclude that the actual value of tidal dissipation is very different form what has been assumed here. Moreover, our approach can also be extended to the case of fully convective late type M-dwarfs, with some words of caution. First, it is expected that because they are fully convective, tidal dissipation in those stars should be weak \citep{Wu2005,Mathis2015}. Second, the absence of a tachoclyne may also affect how magnetic field is generated in those stars, which could imply a braking-law that is not Skumanich-like. Tying age to rotation for late-type M dwarfs is very challenging because both their age and rotation period are difficult to measure, but there is observational evidence that older stars are generally slow rotators for both early-type and late-type M dwarfs \citep{West2015}. Overlooking the complexity of the saturated regime for fast rotators, the magnetic braking law that we used here could also apply to late-type M-dwarfs, using an appropriate value of $\alpha_{\rm mb}$ \citep{Reiners2012}. 

The orbital and rotation periods corresponding to the stationary state are dependant on the value of the assumed $Q'$ and $\subrm{\alpha}{mb}$. The lower the $Q'$, the lower the star's stationary spin period and therefore the faster the loss of angular momentum and orbital decay of the planet in the stationary state. But whereas $\tau_a$ is directly proportional to $Q'$ when the stationary state is unstable, it does not decrease as fast with decreasing $Q'$ when the stationary state is stable. In the same way, $\tau_a$ in the stable stationary state does not scale as $\subrm{\alpha}{mb}$, because the stationary spin has a non-linear dependance with this parameter through Eq.~\ref{eqstat}. This is illustrated on Fig.~\ref{fig:perlim}, that shows the value of the minimum orbital period allowing $\tau_a \geq 5$~Gyrs for a 80~M$_{\rm Jup}$ companion orbiting either an F5, G5 or M5-type star, as a function of $Q'$ and $\gamma= \subrm{\alpha}{mb}/1.5 \times 10^{-14}$~yr. 

Given the wide range of possible values for $\subrm{\alpha}{mb}$ and $Q'$ considered here, we can safely conclude that orbital decay for massive companions can be neglected for orbital periods greater than 10 days.

We also see that if $Q'\lesssim10^7$ and $\gamma\gtrsim0.6$ for G-type stars, and if $Q'\gtrsim4\times10^7$ and $\gamma\lesssim0.4$ for F-type stars, the lack of massive companions on orbital periods shorter than 5 days around G-type stars compared to F-type stars could indeed result from the combined effect of tidal interaction and magnetic braking. This is consistent with the fact that stars with a more extended convective zone would have a greater tidal dissipation efficiency and stronger magnetic braking, as stated in \citet{Bouchy2011}. Due to the strong dependance of the tidal torque with the stellar radius, massive companions can also survive on very close orbits around late-M type stars, even if there magnetic braking is efficient.

\section{Conclusions}\label{sec:conclusion}
We have shown that M-type stars, either fully convective or not are capable of harbouring massive companions on close orbits for extended periods of time. The lack of detection of massive companions orbiting convective stars on short orbit can not only be due to the combined effect of tidal dissipation and magnetic braking on the orbital decay. However, for massive companions on orbital periods shorter than 5 days, those effect may very well explain the lack of systems detected around a G-type host, in agreement with \citet{Bouchy2011} and \citet{Guillot2014}. Nevertheless, assuming reasonable values for the magnetic braking and tidal dissipation factors, we would not expect those effects to produce a difference in the proportion of G and F-type hosts on orbital periods greater than $\sim 6$~days. Extending the range of values for those parameters, we show that tidal decay can be safely neglected for massive companions on orbit greater than $\sim10$~days. Unfortunately, the small number of transiting BD is not providing us with a reliable statistics yet. Moreover, caution must be taken, since we cannot rule out the effect of selection biases that may give a lesser priority to the full characterisation of systems displaying rather deep transits. In the near future, the estimation of the completeness and observational biases of CoRoT and Kepler down to very low mass host stars will be available. It will establish the actual relative frequency of BD around F and latter-type stars. If there is a lack of close-in BD around late M-type stars compared to F-type stars, this could be the signature of an inefficient formation process and would help constraining formation and evolution theories of brown dwarfs and massive extrasolar planets. 

\begin{acknowledgements}
The authors are grateful to the referee, S. Mathis, for valuable comments that improved the manuscript. CD wishes to thank J.P. Marques and F. Baudin for helpful discussions and for providing the CESAM evolutionary tracks. RFD carried out this work within the frame of the National Center for Competence in Research "PlanetS" supported by the Swiss National Science Foundation (SNSF). CD acknowledges funding from the ANR (Agence Nationale de la Recherche, France) program IDEE (ANR-12-BS05-0008) "Interaction Des \'Etoiles et des Exoplan\`etes".
\end{acknowledgements}

 \bibliographystyle{aa} 
 \bibliography{BD} 

\begin{thebibliography}{37}
\expandafter\ifx\csname natexlab\endcsname\relax\def\natexlab#1{#1}\fi

\bibitem[{{Baglin} {et~al.}(2006){Baglin}, {Auvergne}, {Boisnard}, {Lam-Trong},
  {Barge}, {Catala}, {Deleuil}, {Michel}, \& {Weiss}}]{Baglin2006}
{Baglin}, A., {Auvergne}, M., {Boisnard}, L., {et~al.} 2006, in COSPAR Meeting,
  Vol.~36, 36th COSPAR Scientific Assembly, 3749

\bibitem[{Barker \& Ogilvie(2009)}]{Barker2009}
Barker, A.~J. \& Ogilvie, G.~I. 2009, \mnras, 395, 2268

\bibitem[{{Bolmont} {et~al.}(2012){Bolmont}, {Raymond}, {Leconte}, \&
  {Matt}}]{Bolmont2012}
{Bolmont}, E., {Raymond}, S.~N., {Leconte}, J., \& {Matt}, S.~P. 2012, \aap,
  544, A124

\bibitem[{{Borucki} {et~al.}(2010){Borucki}, {Koch}, {Basri}, {Batalha},
  {Brown}, {Caldwell}, {Caldwell}, {Christensen-Dalsgaard}, {Cochran},
  {DeVore}, {Dunham}, {Dupree}, {Gautier}, {Geary}, {Gilliland}, {Gould},
  {Howell}, {Jenkins}, {Kondo}, {Latham}, {Marcy}, {Meibom}, {Kjeldsen},
  {Lissauer}, {Monet}, {Morrison}, {Sasselov}, {Tarter}, {Boss}, {Brownlee},
  {Owen}, {Buzasi}, {Charbonneau}, {Doyle}, {Fortney}, {Ford}, {Holman},
  {Seager}, {Steffen}, {Welsh}, {Rowe}, {Anderson}, {Buchhave}, {Ciardi},
  {Walkowicz}, {Sherry}, {Horch}, {Isaacson}, {Everett}, {Fischer}, {Torres},
  {Johnson}, {Endl}, {MacQueen}, {Bryson}, {Dotson}, {Haas}, {Kolodziejczak},
  {Van Cleve}, {Chandrasekaran}, {Twicken}, {Quintana}, {Clarke}, {Allen},
  {Li}, {Wu}, {Tenenbaum}, {Verner}, {Bruhweiler}, {Barnes}, \&
  {Prsa}}]{2010Sci...327..977B}
{Borucki}, W.~J., {Koch}, D., {Basri}, G., {et~al.} 2010, Science, 327, 977

\bibitem[{{Bouchy} {et~al.}(2011){Bouchy}, {Deleuil}, {Guillot}, {Aigrain},
  {Carone}, {Cochran}, {Almenara}, {Alonso}, {Auvergne}, {Baglin}, {Barge},
  {Bonomo}, {Bord{\'e}}, {Csizmadia}, {de Bondt}, {Deeg}, {D{\'{\i}}az},
  {Dvorak}, {Endl}, {Erikson}, {Ferraz-Mello}, {Fridlund}, {Gandolfi},
  {Gazzano}, {Gibson}, {Gillon}, {Guenther}, {Hatzes}, {Havel}, {H{\'e}brard},
  {Jorda}, {L{\'e}ger}, {Lovis}, {Llebaria}, {Lammer}, {MacQueen}, {Mazeh},
  {Moutou}, {Ofir}, {Ollivier}, {Parviainen}, {P{\"a}tzold}, {Queloz}, {Rauer},
  {Rouan}, {Santerne}, {Schneider}, {Tingley}, \& {Wuchterl}}]{Bouchy2011}
{Bouchy}, F., {Deleuil}, M., {Guillot}, T., {et~al.} 2011, \aap, 525, A68

\bibitem[{{Burrows} {et~al.}(1997){Burrows}, {Marley}, {Hubbard}, {Lunine},
  {Guillot}, {Saumon}, {Freedman}, {Sudarsky}, \& {Sharp}}]{Burrows1997}
{Burrows}, A., {Marley}, M., {Hubbard}, W.~B., {et~al.} 1997, \apj, 491, 856

\bibitem[{{Chabrier} \& {Baraffe}(2000)}]{Chabrier2000}
{Chabrier}, G. \& {Baraffe}, I. 2000, \araa, 38, 337

\bibitem[{{Chabrier} {et~al.}(2014){Chabrier}, {Johansen}, {Janson}, \&
  {Rafikov}}]{Chabrier2014}
{Chabrier}, G., {Johansen}, A., {Janson}, M., \& {Rafikov}, R. 2014, Protostars
  and Planets VI, 619

\bibitem[{{Cox}(2000)}]{Allen}
{Cox}, A.~N. 2000, {Allen's astrophysical quantities}

\bibitem[{{Csizmadia} {et~al.}(2015){Csizmadia}, {Hatzes}, {Gandolfi},
  {Deleuil}, {Bouchy}, {Fridlund}, {Szabados}, {Parviainen}, {Cabrera},
  {Aigrain}, {Alonso}, {Almenara}, {Baglin}, {Bord{\'e}}, {Bonomo}, {Deeg},
  {D{\'{\i}}az}, {Erikson}, {Ferraz-Mello}, {Tadeu dos Santos}, {Guenther},
  {Guillot}, {Grziwa}, {H{\'e}brard}, {Klagyivik}, {Ollivier}, {P{\"a}tzold},
  {Rauer}, {Rouan}, {Santerne}, {Schneider}, {Mazeh}, {Wuchterl}, {Carpano}, \&
  {Ofir}}]{Csizmadia2015}
{Csizmadia}, S., {Hatzes}, A., {Gandolfi}, D., {et~al.} 2015, \aap, 584, A13

\bibitem[{{Damiani} \& {Lanza}(2015)}]{Damiani2015}
{Damiani}, C. \& {Lanza}, A.~F. 2015, \aap, 574, A39

\bibitem[{{Deleuil} {et~al.}(2008){Deleuil}, {Deeg}, {Alonso}, {Bouchy},
  {Rouan}, {Auvergne}, {Baglin}, {Aigrain}, {Almenara}, {Barbieri}, {Barge},
  {Bruntt}, {Bord{\'e}}, {Collier Cameron}, {Csizmadia}, {de La Reza},
  {Dvorak}, {Erikson}, {Fridlund}, {Gandolfi}, {Gillon}, {Guenther}, {Guillot},
  {Hatzes}, {H{\'e}brard}, {Jorda}, {Lammer}, {L{\'e}ger}, {Llebaria},
  {Loeillet}, {Mayor}, {Mazeh}, {Moutou}, {Ollivier}, {P{\"a}tzold}, {Pont},
  {Queloz}, {Rauer}, {Schneider}, {Shporer}, {Wuchterl}, \&
  {Zucker}}]{Deleuil2008}
{Deleuil}, M., {Deeg}, H.~J., {Alonso}, R., {et~al.} 2008, \aap, 491, 889

\bibitem[{{Dobbs-Dixon} {et~al.}(2004){Dobbs-Dixon}, {Lin}, \&
  {Mardling}}]{Dobbs-Dixon2004}
{Dobbs-Dixon}, I., {Lin}, D.~N.~C., \& {Mardling}, R.~A. 2004, \apj, 610, 464

\bibitem[{{Ferraz-Mello} {et~al.}(2015){Ferraz-Mello}, {Tadeu dos Santos},
  {Folonier}, {Czismadia}, {do Nascimento}, \& {P{\"a}tzold}}]{Ferraz2015}
{Ferraz-Mello}, S., {Tadeu dos Santos}, M., {Folonier}, H., {et~al.} 2015,
  \apj, 807, 78

\bibitem[{{Grether} \& {Lineweaver}(2006)}]{Grether2006}
{Grether}, D. \& {Lineweaver}, C.~H. 2006, \apj, 640, 1051

\bibitem[{{Guillot} {et~al.}(2014){Guillot}, {Lin}, {Morel}, {Havel}, \&
  {Parmentier}}]{Guillot2014}
{Guillot}, T., {Lin}, D.~N.~C., {Morel}, P., {Havel}, M., \& {Parmentier}, V.
  2014, in EAS Publications Series, Vol.~65, EAS Publications Series, 327--336

\bibitem[{{Hansen}(2010)}]{Hansen2010}
{Hansen}, B.~M.~S. 2010, \apj, 723, 285

\bibitem[{Hut(1980)}]{1980A&A....92..167H}
Hut, P. 1980, \aap, 92, 167

\bibitem[{{Jackson} {et~al.}(2009){Jackson}, {Barnes}, \&
  {Greenberg}}]{Jackson2009}
{Jackson}, B., {Barnes}, R., \& {Greenberg}, R. 2009, \apj, 698, 1357

\bibitem[{Kawaler(1988)}]{Kawaler1988}
Kawaler, S.~D. 1988, \apj, 333, 236

\bibitem[{{Luhman}(2012)}]{Luhman2012}
{Luhman}, K.~L. 2012, \araa, 50, 65

\bibitem[{{Ma} \& {Ge}(2014)}]{Ma2014}
{Ma}, B. \& {Ge}, J. 2014, \mnras, 439, 2781

\bibitem[{{Marcy} \& {Butler}(2000)}]{Marcy2000}
{Marcy}, G.~W. \& {Butler}, R.~P. 2000, \pasp, 112, 137

\bibitem[{{Mathis}(2015{\natexlab{a}})}]{Mathis2015sf2a}
{Mathis}, S. 2015{\natexlab{a}}, in SF2A-2015: Proceedings of the Annual
  meeting of the French Society of Astronomy and Astrophysics. Eds.: F.
  Martins, S. Boissier, V. Buat, L. Cambr{\'e}sy, P. Petit, pp.401-405, ed.
  F.~{Martins}, S.~{Boissier}, V.~{Buat}, L.~{Cambr{\'e}sy}, \& P.~{Petit},
  401--405

\bibitem[{{Mathis}(2015{\natexlab{b}})}]{Mathis2015}
{Mathis}, S. 2015{\natexlab{b}}, \aap, 580, L3

\bibitem[{{Mathis} \& {Remus}(2013)}]{Mathis2013}
{Mathis}, S. \& {Remus}, F. 2013, in Lecture Notes in Physics, Berlin Springer
  Verlag, Vol. 857, Lecture Notes in Physics, Berlin Springer Verlag, ed. J.-P.
  {Rozelot} \& C.~. {Neiner}, 111--147

\bibitem[{Matt {et~al.}(2012)Matt, MacGregor, Pinsonneault, \&
  Greene}]{Matt2012}
Matt, S.~P., MacGregor, K.~B., Pinsonneault, M.~H., \& Greene, T.~P. 2012,
  \apj, 754, L26

\bibitem[{{Morel}(1997)}]{1997A&AS..124..597M}
{Morel}, P. 1997, \aaps, 124, 597

\bibitem[{{Morel} \& {Lebreton}(2008)}]{2008Ap&SS.316...61M}
{Morel}, P. \& {Lebreton}, Y. 2008, \apss, 316, 61

\bibitem[{{Ogilvie}(2014)}]{Ogilvie2014}
{Ogilvie}, G.~I. 2014, \araa, 52, 171

\bibitem[{Ogilvie \& Lin(2007)}]{Ogilvie2007}
Ogilvie, G.~I. \& Lin, D. N.~C. 2007, \apj, 661, 1180

\bibitem[{Reiners \& Mohanty(2012)}]{Reiners2012}
Reiners, A. \& Mohanty, S. 2012, The Astrophysical Journal, 746, 43

\bibitem[{{R{\'e}ville} {et~al.}(2015){R{\'e}ville}, {Brun}, {Matt},
  {Strugarek}, \& {Pinto}}]{Reville2015}
{R{\'e}ville}, V., {Brun}, A.~S., {Matt}, S.~P., {Strugarek}, A., \& {Pinto},
  R.~F. 2015, \apj, 798, 116

\bibitem[{{Sahlmann} {et~al.}(2011){Sahlmann}, {S{\'e}gransan}, {Queloz},
  {Udry}, {Santos}, {Marmier}, {Mayor}, {Naef}, {Pepe}, \&
  {Zucker}}]{Sahlmann2011}
{Sahlmann}, J., {S{\'e}gransan}, D., {Queloz}, D., {et~al.} 2011, \aap, 525,
  A95

\bibitem[{{Santerne} {et~al.}(2015){Santerne}, {Moutou}, {Tsantaki}, {Bouchy},
  {H{\'e}brard}, {Adibekyan}, {Almenara}, {Amard}, {Barros}, {Boisse},
  {Bonomo}, {Bruno}, {Courcol}, {Deleuil}, {Demangeon}, {D{\'{\i}}az},
  {Guillot}, {Havel}, {Montagnier}, {Rajpurohit}, {Rey}, \&
  {Santos}}]{2015arXiv151100643S}
{Santerne}, A., {Moutou}, C., {Tsantaki}, M., {et~al.} 2015, ArXiv e-prints
  [\eprint[arXiv]{1511.00643}]

\bibitem[{{West} {et~al.}(2015){West}, {Weisenburger}, {Irwin},
  {Berta-Thompson}, {Charbonneau}, {Dittmann}, \& {Pineda}}]{West2015}
{West}, A.~A., {Weisenburger}, K.~L., {Irwin}, J., {et~al.} 2015, \apj, 812, 3

\bibitem[{{Wu}(2005)}]{Wu2005}
{Wu}, Y. 2005, \apj, 635, 688

\end{thebibliography}

\end{document}